\documentclass[aip, jap, reprint]{revtex4-1}

\usepackage{graphicx}
\draft 

\begin{document}

\title{Exchange coupling and magnetoresistance in CoFe/NiCu/CoFe spin-valves near the Curie point of the spacer} 
\author{S. Andersson and V. Korenivski}

\affiliation{Nanostructure Physics, Royal Institute of Technology, SE-106 91 Stockholm, Sweden}

\date{\today}

\begin{abstract}
Thermal control of exchange coupling between two strongly ferromagnetic layers through a weakly ferromagnetic Ni-Cu spacer and the associated magnetoresistance is investigated. The spacer, having a Curie point slightly above room temperature, can be cycled between its paramagnetic and ferromagnetic states by varying the temperature externally or using joule heating. It is shown that the giant magnetoresistance vanishes due to a strong reduction of the mean free path in the spacer at  above $\sim$30\% Ni concentration --- before the onset of ferromagnetism. Finally, a device is proposed and demonstrated which combines thermally controlled exchange coupling and large magnetoresistance by separating the switching and the read out elements.

\end{abstract}

\maketitle 

\section{Introduction}
Thermal control in spintronic devices and MRAM applications has in recent years been of great interest due to the associated increase of stability and decrease in power consumption \cite{ref:Prejbeanu, ref:Beech, ref:Daughton}.  Recently, thermally excited oscillations in nanocontacts, reaching frequencies of the order of GHz, have been predicted \cite{ref:Kadigrobov}. In this model a ferromagnetic (FM) film is separated from a small FM grain by a point contact having a diameter of a few nanometers. Due to the high current densities reached in such a small area, the FM region within the point contact reaches very high temperatures. When the local temperature is higher than the FM Curie point the exchange coupling through the point contact becomes vanishingly small. 

The model is based on two premises. First, a thermally controlled exchange coupling between two FM regions. Second, an increase in resistance when the FM regions decouple. The first criterion can be met in a metallic system with two strong ferromagnets separated by a weakly FM spacer. If this can be combined with a large change in resistance both criteria would be met. To date, the largest changes of resistance obtained in an all metallic structure have been from giant magnetoresistance \cite{ref:Binasch, ref:Baibich} (GMR).

In this work we investigate the possibility of decoupling two strong ferromagnets separated by a weakly ferromagnetic Ni-Cu alloy. To verify if the Ni-Cu alloy, in its paramagnetic phase, can be used as a GMR spacer we study the effects of adding nickel to a copper spacer in a spin-valve structure on the interlayer exchange coupling and GMR. Finally, we design and implement an improved thermionic spin-valve structure, in which the switching and the read out layers are separated.

\section{Experimental details}
All films were deposited on thermally oxidized Si substrates using magnetron sputtering at a base pressure better than $5\cdot10^{-8}$ Torr. The argon pressure during sputtering was kept at 3 mTorr. To demonstrate thermally controlled exchange coupling samples with structure Cu(90)/Ni$_{80}$Fe$_{20}$(8)/Co$_{90}$Fe$_{10}$(2)/ Ni$_{x}$Cu$_{1-x}$(t)/Co$_{90}$Fe$_{10}$(5)/Ta(10) (thickness in nanometers) were deposited. Three different thicknesses, $t = 10, 20, 30$ nm,  of the weakly ferromagnetic Ni-Cu alloy were used for studying the interlayer exchange in the system. Variation in $x$ was obtained by cosputtering Ni and Cu onto Si substrates that had been cut into 90 x 10 mm strips. In this way a compositional gradient was created along the Si strips ranging from $x=0.2$ to $x=0.9$. By cutting the strips into smaller pieces a series of samples with different Curie temperatures were obtained. 

In order to perform magnetic characterization the samples were placed in a looptracer equipped with a thin-film heater. The temperature was controlled through a feed back loop using a type-T thermocouple in close contact with the samples. Further investigations of the switching behavior at room temperature were performed using a vibrating sample magnetometer (VSM).

To measure the effect of Ni-Cu alloying on GMR, samples with structure Ni$_{80}$Fe$_{20}$(4) / Co$_{90}$Fe$_{10}$(1) / Ni$_{x}$Cu$_{1-x}$(3.5) / Co$_{90}$Fe$_{10}$(2) were deposited. The Ni-Cu alloys were cosputtered from Ni and Cu targets such that the stoichiometry, $x$, could be varied by controlling the relative difference in sputtering rates. 

To measure current in plane (CIP) GMR, thin Al wires were bonded to the top of the samples. Before electrical measurements
the separate switching of the NiFe/CoFe bi-layer and CoFe top layer was confirmed using a magnetometer.

\section{Results and Discussion}
A Ni-Cu alloy was chosen for the weakly FM spacer because of the well known dependence of Curie point on the Ni concentration \cite{ref:Hicks, ref:Dutta, ref:Sousa}. The Ni-Cu spacer is used to separate a magnetically softer NiFe/CoFe bi-layer from a magnetically harder CoFe layer. Here NiFe and CoFe stand for Ni$_{80}$Fe$_{20}$ and Co$_{90}$Fe$_{10}$, respectively. By cosputtering Ni and Cu, a series of samples with different Curie temperatures were obtained. 

\begin{figure}[htbp]
   \centering
   \includegraphics[width=3in]{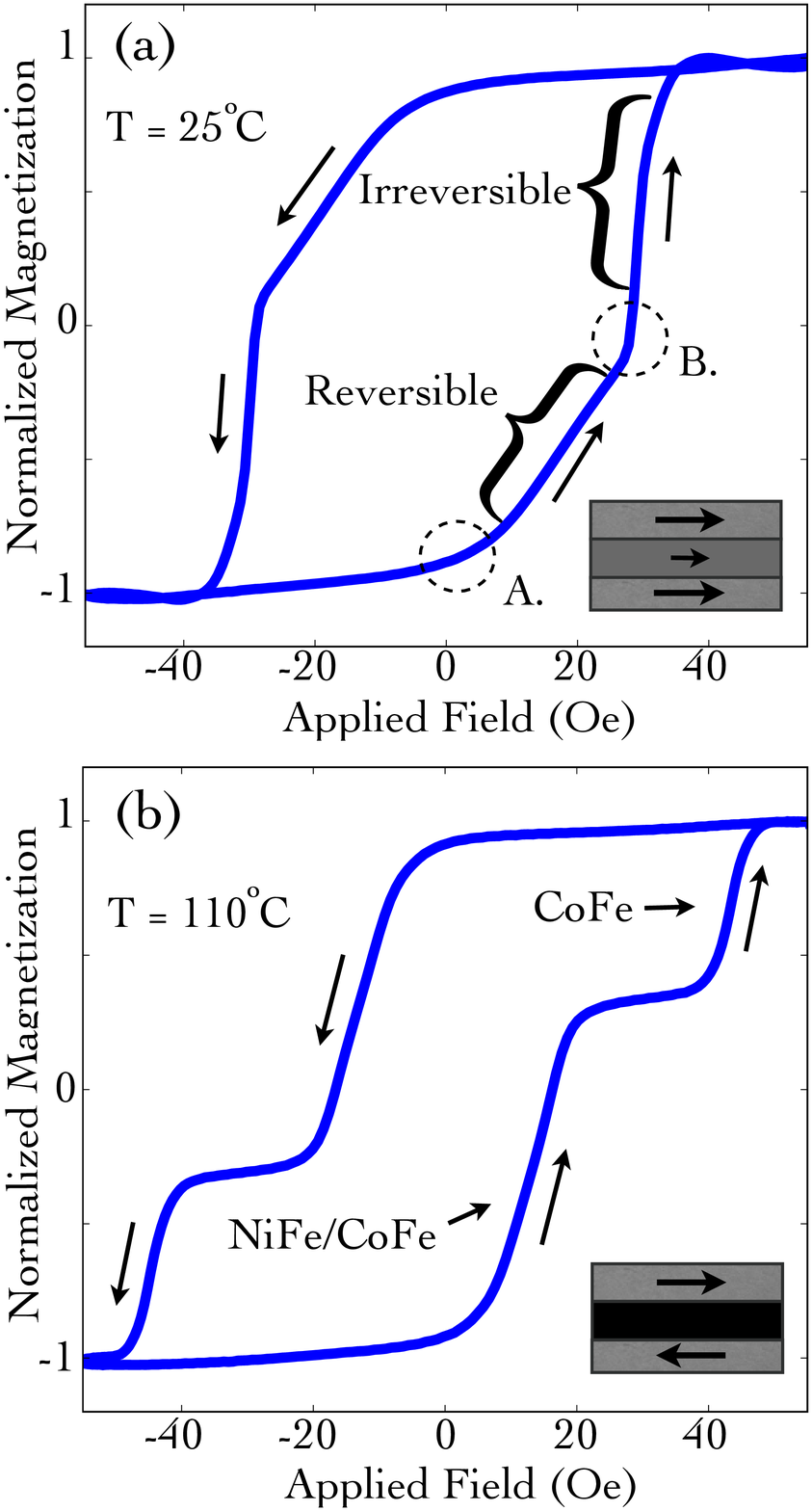}
   \caption{Normalized magnetization versus applied magnetic field for a sample structure NiFe/CoFe/Ni$_{x}$Cu$_{1-x}$(30)/ CoFe, x $\approx$ 0.7 at (a) room temperature and (b) 110$^{\circ}$C. Upon heating the Ni-Cu alloy goes through a ferromagnetic to paramagnetic phase transition and the NiFe/CoFe bi-layer decouples from the CoFe layer.}
   \label{fig:Fig1}
\end{figure}

\subsection{Thermally controlled interlayer exchange coupling}
Fig. 1 (a) shows the magnetization loop for a sample with a 30 nm thick Ni-Cu layer having a Curie point just above room temperature, $x \approx 0.7$. The shape of the curve indicates that the strongly FM layers are weakly exchange coupled through the Ni-Cu alloy. Two distinct regions can be seen --- similar to the magnetic state of a spring-magnet \cite{ref:Davies}. At point $A$ in Fig. 1 (a) the magnetic moment of the soft NiFe/CoFe layer starts to rotate in the external magnetic field. This is a reversible rotation due to the exchange coupling through the Ni-Cu alloy to the harder CoFe layer. The reversible switching continues until point $B$ is reached. By comparing the magnitude of the magnetization at 25 Oe (point B) in Fig. 1 (a) with that in Fig. 1 (b) in which the NiFe/CoFe layer is heated to 110$^{\circ}$C and thereby decoupled from the CoFe layer, it can be seen that at room temperature the NiFe/CoFe layer has not yet finished rotating 180$^{\circ}$ when the hard layer switches. At point $B$ the torque on the CoFe layer is too strong for it to remain in position and an irreversible rotation of all layers follows. This behavior was confirmed by VSM measurements of the same sample at room temperature. Starting at a positive field, high enough so that all the moments were aligned, the magnetization was measured while the external field was swept to -20 Oe and then back again. After the field reversal at -20 Oe the magnetization backtracks the values measured before the reversal. The same behavior was seen for field reversals up to -25 Oe indicating that the rotation is reversible up to this point. For reversal fields any higher than this, the magnetization does not backtrack the values measured before the field reversal. This confirms that the switching behavior is irreversible for fields higher than $\pm$25 Oe.

Fig. 1 (b) shows the same sample at $110^{\circ}$C. The Curie point of the spacer has been reached and the soft and hard FM layers are essentially exchange decoupled as evidenced by the two distinct magnetization transitions at approximately 15 and 45 Oe. As the temperature is reduced to room temperature the two magnetization transitions shift towards each other and the sharp magnetization loop becomes significantly skewed. At room temperature the curve shape is back to the one shown in Fig. 1 (a). This thermally controlled interlayer exchange coupling is perfectly reversible on thermal cycling within the given temperature range.

For samples with the Curie point around room temperature the spacer had to be 20 or 30 nm thick in order to completely diminsh the exchange coupling through the spacer. An explanation for this could be that the alloy is not homogenous after cosputtering at room temperature but contains regions with different Curie points. If the spacer is too thin, these regions could extend to the alloy interfaces and couple the two CoFe films. Another possible explanation is that the alloying is homogenous but the two strong ferromagnets are coupled by exchange interactions through the spacer even at temperatures above the Curie point. It has been indicated that exchange can propagate through paramagnetic regions on length scales of several nanometers \cite{ref:Hernando}, which is believed to be due to enhancement of magnetic order in thin layers caused by the proximity effect of a strong ferromagnet.

\begin{figure}[htbp]
   \centering
   \includegraphics[width=3in]{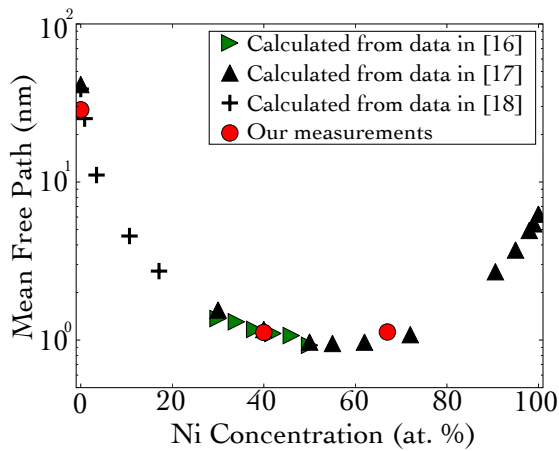}
   \caption{Electron mean free path in bulk Ni-Cu for different Ni concentrations. The data points have been calculated using the Drude model from published data on resistivity measurements at 300, 273 and 250K \cite{ref:Houghton, ref:Ahmad, ref:Schroeder}.}
   \label{fig:Fig2}
\end{figure}

\begin{figure}[htbp]
   \centering
   \includegraphics[width=3in]{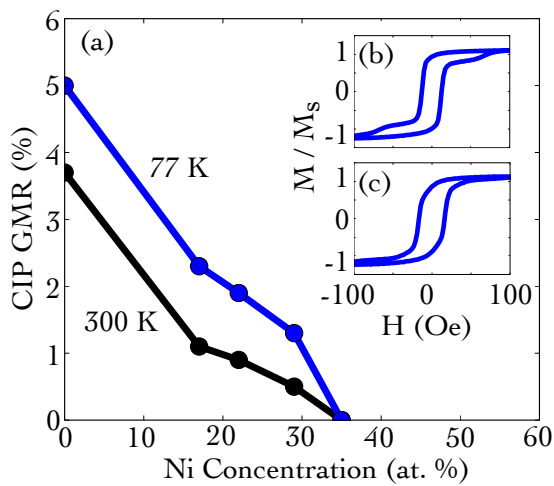}
   \caption{(a) Current in plane (CIP) giant magnetoresistance (GMR) versus Ni concentration in a NiFe(4)/CoFe(1)/Ni-Cu(3.5)/CoFe(2) spin-valve. The inset shows the normalized magnetization versus applied field for a spin-valve with (b) 29 at.\% Ni in the spacer and (c) 35 at. \% Ni in the spacer.}
   \label{fig:Fig3}
\end{figure}

\subsection{CoFe/Ni-Cu/CoFe spin-valve}
To understand the CIP GMR in the above NiFe/CoFe/Ni-Cu/CoFe system, we have to consider the mean free path in the Ni-Cu spacer. When the spacer thickness is much larger than the mean free path the CIP GMR signal vanishes  as exp$(-t_{NiCu}/\lambda)$ \cite{ref:Barthelemy}. Here $\lambda$ is the electron mean free path in the spacer material and $t_{NiCu}$ is the spacer thickness. In order to obtain a high GMR signal the thickness of the spacer should be comparable or smaller than $\lambda$. Fig. 2 shows $\lambda$ for different Ni concentrations from our measurements as well as calculated from published experimental data using the Drude model. The mean free path decreases quickly with increasing Ni concentration. In the interesting range of Ni concentrations between 40\% and 70\% where the alloy is weakly ferromagnetic \cite{ref:Hicks}, $\lambda$ is down to $\sim$1 nm. As was detailed in the previous section, our spacers with the Curie point close to room temperature have a minimum thickness of 20 nm in order to completely diminish the interlayer exchange coupling. 
Assuming the thickness of the weakly ferromagnetic spacer can be reduced by further material optimization, we next examine how thin a spacer would still provide a measurable GMR signal. The thinnest possible spacer to avoid any significant RKKY coupling is $\sim$3 nm \cite{ref:Parkin}. We therefore choose this thickness and evaluate the CIP GMR versus Ni concentration.

We have fabricated samples with structure NiFe(4)/CoFe(1)/Ni-Cu(3.5)/CoFe(2). CIP GMR measurements for samples with different Ni concentrations are shown in Fig. 3 (a). With pure Cu in the spacer a signal of 3.7 \% is measured at room temperature (5\% at 77 K). When Ni is added to the spacer the signal drops caused by a decrease in $\lambda$. At 29 at. \% Ni the GMR has decreased by more than a factor of three, which we attribute to a sharp decrease of the mean free path on alloying. Fig. 3 (b) shows the magnetization versus applied field for this sample. It can be seen that the switching is still separate at this concentration. At 35 at. \% Ni the magnetic layers couple and the switching is no longer separate, which is shown in Fig. 3 (c). It can thus be concluded that even for a very thin Cu-Ni spacer the CIP GMR signal essentially vanishes for 29 at. \% Ni concentration, where the layers are still decoupled magnetically. The final decrease to zero GMR at 35 at. \% Ni is due to coupling through the spacer and not to a decrease in $\lambda$.

It will be informative to point out that by using current perpendicular to the plane GMR instead of CIP the limiting length scale would be the spin diffusion length  \cite{ref:Bass}, $l_{sf}$, and not $\lambda$. However, published experimental data indicate that $l_{sf}$ decrease with the same rate as $\lambda$ and has been measured to be 7.5 nm at 5$^{\circ}$K in an alloy with 22.7\% Ni \cite{ref:Hsu}. This is still below the minimum thickness in our case of 20 nm needed to achieve reliable interlayer decoupling.

\begin{figure}[htbp]
   \centering
   \includegraphics[width=3in]{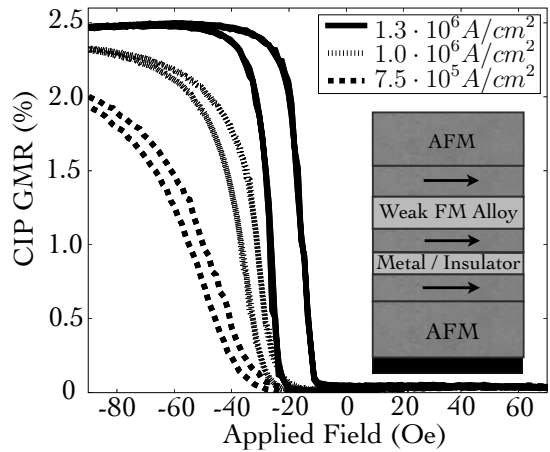}
   \caption{CIP GMR for three different current densities versus applied field measured on a sample structure NiFe(3)/MnIr(15)/CoFe(4)/Cu(3.5)/CoFe(4)/ NiFe(6)/CoFe(2)/NiCu(20)/CoFe(2)/NiFe(3)/ MnIr(12)/ Ta(5). The inset shows a schematic of a device in which thermally controlled exchange coupling is separated from a spin-valve read out. Either giant magnetoresistance or tunnelling magnetoresistance can be used for read out.}
   \label{fig:Fig4}
\end{figure}

To overcome the above limitations and use thermally controlled interlayer exchange coupling together with large magnetoresistance we propose a new design in which the GMR read out layer and the weakly FM spacer are separated. A schematic is shown in the inset to Fig. 4. An antiferromagnet (AFM) is used to exchange bias a FM film which works as a reference layer in the spin-valve. The spin-valve uses a metallic spacer for GMR or an insulator for tunneling magnetoresistance. The spin-valve spacer is only used for read out and can be optimized to give the highest possible magnetoresistance signal. The top layer of the spin-valve is exchange coupled through a weakly FM alloy to a top pinned FM. At high temperatures the coupling through the weakly FM alloy is negligible and the top layer of the spin-valve is free to rotate. However, at low temperatures it will be exchange coupled to the top pinned FM. This results in full flexibility when choosing the composition and Curie point of the weakly ferromagnetic alloy while at the same time making it possible to achieve high magnetoresistance.

To demonstrate this new structure we have deposited samples of NiFe(3)/MnIr(15)/CoFe(4)/ Cu(3.5)/CoFe(4)/NiFe(6)/ CoFe(2)/NiCu(20)/CoFe(2)/NiFe(3)/MnIr(12)/Ta(5) with 70 at. \% Ni in the spacer ($T_{C}\approx 100^{\circ}$C). The samples were patterned using photolithography into strips 50 $\mu$m wide and 1 mm long and then bonded at the edges with aluminum wire. The resulting CIP GMR signal for different current densities is shown in FIG. 4. When the current density is increased and the temperature of the device is correspondingly raised due to joule heating, the top layer of the spin-valve decouples from the top pinned CoFe/NiFe bi-layer producing a strong GMR signal. For a current density of $1.3\cdot 10^{6}$ A/cm$^{2}$, $T>T_{C}\approx 100^{\circ}$C, the exchange decoupling is complete and the full CIP GMR signal of 2.5\% for this structure is obtained.

\section{Conclusion}
Thermally controlled exchange coupling between two strongly FM films separated by a weakly FM Ni-Cu spacer is demonstrated. At temperatures higher than the Curie point of the spacer the FM films are decoupled. At lower temperatures the switching behavior can be separated into two regions --- reversible and irreversible.

In a CoFe/Ni-Cu/CoFe spin-valve the CIP GMR signal vanishes due to a sharp reduction of the mean free path on alloying for Ni concentrations above $\sim$30 at. \%.  A new design is proposed and demonstrated, combining thermally controlled exchange coupling and large magnetoresistance, which may prove useful for applications in current controlled magneto-resistive oscillators.

\begin{acknowledgments}
This work was supported by EU-FP7-FET-STELE.
\end{acknowledgments}


\begin{thebibliography}{1}

\bibitem{ref:Prejbeanu}
I. L. Prejbeanu, M. Kerekes, R. C. Sousa, H. Sibuet, O. Redon, B. Dieny and J. P. Nozi\'eres, J. Phys. Condens. Matter \textbf{19}, 165218 (2007).

\bibitem{ref:Beech}
R. S. Beech, J. A. Andersson, A. V. Pohm and J. M. Daughton, J. Appl. Phys. \textbf{87}, 6403 (2000).

\bibitem{ref:Daughton}
J. M. Daughton and A. V. Pohm, J. Appl. Phys. \textbf{93}, 7304 (2003).

\bibitem{ref:Kadigrobov}
A. Kadigrobov, S. I. Kulinich, R. I. Shekhter, M. Jonson and V. Korenivski, Phys. Rev. B \textbf{74}, 195307 (2006).

\bibitem{ref:Binasch}
G. Binasch, P. Gr\"unberg, F. Saurenbach and W. Zinn, Phys. Rev. B \textbf{39}, 4828 (1989).

\bibitem{ref:Baibich}
M. N. Baibich, J. M. Broto, A. Fert, F. Nguyen Van Dau and F. Petroff, Phys. Rev. Lett. \textbf{61}, 2472 (1988).

\bibitem{ref:Hicks}
T. J. Hicks, B. Rainford, J. S. Kouvel and G. G. Low, Phys. Rev. Lett. \textbf{22}, 531 (1969).

\bibitem{ref:Dutta}
S. K. Dutta Roy and A. V. Subrahmanyam, Phys. Rev. Lett. \textbf{177}, 1133 (1969).

\bibitem{ref:Sousa}
J. B. Sousa, M. R. Chaves, M. F. Pinheiro and R. S. Pinto, J. Low. Temp. Phys. \textbf{18}, 125 (1975).

\bibitem{ref:Davies}
J. E. Davies, O. Hellwig, E. E. Fullerton, J. S. Jiang, S. D. Bader, G. T. Zim\'anyi and K. Liu, Appl. Phys. Lett. \textbf{86}, 262503 (2005).

\bibitem{ref:Hernando}
A. Hernando and T. Kulik, Phys. Rev. B \textbf{49}, 7064 (1994).

\bibitem{ref:Barthelemy}
A. Barth\'el\'emy and A. Fert, Phys. Rev. B \textbf{43}, 13124 (1991).

\bibitem{ref:Parkin}
S. S. P. Parkin, R. Bhadra and K. P. Roche, Phys. Rev. Lett. \textbf{66}, 2152 (1991).

\bibitem{ref:Bass}
J. Bass and W. P. Pratt Jr., J. Phys. Condens. Matter \textbf{19}, 183201 (2007).

\bibitem{ref:Hsu}
S. Y. Hsu, P. Holody, R. Loloee, J. M. Rittner, W. P. Pratt Jr. and P. A. Schroeder, Phys. Rev. B \textbf{54}, 9027 (1996).

\bibitem{ref:Houghton}
R. W. Houghton, M. P. Sarachik and J. S. Kouvel, Phys. Rev. Lett. \textbf{25}, 238 (1970).

\bibitem{ref:Ahmad}
H. M. Ahmad and D. Greig, J. Phys. (Paris). \textbf{35}, C4-223 (1974).

\bibitem{ref:Schroeder}
P. A. Schroeder, R. Wolf and J. A. Woollam, Phys. Rev. \textbf{138}, A105 (1965).

\end{thebibliography}
\end{document}